# МЕТОДЫ РЕГУЛИРОВАНИЯ БЮДЖЕТНЫХ ФИНАНСОВЫХ РИСКОВ В РАМКАХ РЕАЛИЗАЦИИ МЕРОПРИЯТИЙ ПОДДЕРЖКИ ПРЕДПРИНИМАТЕЛЬСТВА*

Елена Георгиевна ДМИТРИК

кандидат экономических наук, доцент кафедры экономики, управления и организации производства, Старооскольский технологический институт им. А.А. Угарова (филиал) Национального исследовательского технологического университета «МИСиС», Старый Оскол, Российская Федерация
dmitrikeg@mail.ru

**Аннотация**

**Предмет.** Основными проблемами, сдерживающими развитие предпринимательства, являются недостаточность собственных финансовых ресурсов, ограниченная возможность привлечения банковских ссуд и бюджетных ассигнований, риски предпринимательской деятельности, а также доступность привлечения финансовой поддержки. Рассмотрена схема бюджетного финансирования мероприятий финансовой поддержки предпринимательства на условиях минимизации финансовых рисков муниципалитета и льготных условий для субъектов предпринимательства.
**Цели.** Выработка научно обоснованных финансовых механизмов финансирования инвестиционных процессов в приоритетных направлениях развития территорий.
**Методология.** Использовались экономико-математические методы, логический анализ и синтез.
**Результаты.** Разработан инструментарий финансирования приоритетных направлений экономики муниципального образования посредством формирования «Фонда инвестиций в микропредприятия» за счет средств муниципальных бюджетов, выпуска долговых муниципальных ценных бумаг и обеспечения части обязательств заемщика по кредиту Фонда государственными (муниципальными) долговыми ценными бумагами. Предложенная финансовая схема позволяет расширить объемы финансирования и обеспечить доступ субъектов предпринимательства к финансовой поддержке на приемлемых и простых условиях как для заемщика, так и для кредитора. Предложены механизмы финансирования инвестиционных предпринимательских проектов на условиях минимальных рисков для бюджета и льготных условиях для субъектов предпринимательства. Представлена методика, позволяющая при прогнозируемом финансовом риске расширить объемы государственного финансирования мероприятий поддержки предпринимательства.
**Выводы.** Предложенная методика регулирования размера кредитной ставки в зависимости от объемов муниципального поручительства и приоритетности вида предпринимательской деятельности позволяет прогнозировать и управлять кредитными рисками так, чтобы риск (убыток) вложения бюджетных средств оставался минимальным или полностью отсутствовал, при этом обеспечивается доступность источников финансирования для субъектов предпринимательства.

Проблемы стимулирования предпринимательской активности и развития частного сектора экономики остаются актуальными для России на протяжении нескольких десятилетий [1, 2]. При научно обоснованном подходе к проблемам стимулирования предпринимательской активности возможно в ближайшей перспективе активизировать процессы продуктивного творчества и добиться увеличения доли инновационного, креативного и ноосферного сегмента экономики, развития малого бизнеса, новых технологий в основных отраслях народного хозяйства [3], что будет способствовать реиндустриализации экономики.

Актуальность проблемы стимулирования предпринимательской активности в условиях трансформации внешней среды и эффективности мер государственной политики в данной области определяется необходимостью и потребностью комплексного исследования и разработки теоретико-методологических положений по оптимизации направлений стимулирования предпринимательства на основе баланса интересов всех субъектов.

Одной из основных проблем, сдерживающих развитие предпринимательства, является недостаточность собственных финансовых ресурсов у субъектов предпринимательской деятельности [4, 5]. Но возможность привлечения банковских ссуд ограничена невысокой доходностью большей части мелкого бизнеса (ниже ставок по кредитам), сложностью их получения (поручительство, залог), рисками предпринимательской деятельности, что в итоге проводит к возникновению просроченной задолженности, снижению эффективности деятельности хозяйствующих субъектов[1]. Возникновение проблемной задолженности обусловлено рисками, которые связаны как с самим заемщиком, так и с внутренними рисками финансовых институтов и рисками внешней среды. Однако в виде результата возможного неблагоприятного исхода их влияние распространяется на всех участников кредитного процесса [6]. Поэтому разработка схем финансирования приоритетных направлений поддержки предпринимательства в рамках программ развития территорий и поиск источников их финансирования на основе реализации принципа сбалансированности интересов бюджетов и хозяйствующих субъектов является задачей, требующей научно обоснованных подходов к ее решению.

Другой не менее важной проблемой является доступность для субъектов предпринимательства государственной финансовой поддержки[2] [7].

В целях расширения доступа субъектов предпринимательской деятельности к финансовым ресурсам реализуются различные программы поддержки предпринимательства как на федеральном, так и на региональном и муниципальном уровнях [8]. Финансирование мероприятий и программ поддержки осуществляется в условиях ограниченности бюджетных ассигнований, что диктует необходимость более жесткого контроля за расходованием бюджетных средств, ужесточения требований к получателям поддержки и в итоге затрудняет доступ субъектов предпринимательства к финансовым ресурсам, провоцирует обострение проблем социально-экономического характера. Данные факторы вызывают необходимость дальнейшего расширения и укрепления финансовой базы и решения ряда проблем, связанных с совершенствованием методов формирования и использования финансовых ресурсов муниципальных образований[3] [9].

Для минимизации бюджетных финансовых рисков и совершенствования инструментов финансовой поддержки предпринимательства мы предлагаем к рассмотрению схему выпуска долговых муниципальных ценных бумаг в виде переводных векселей со сроком погашения «по предъявлению» на объем средств, предусмотренных программами поддержки предпринимательства, где векселедателем и получателем платежа по векселю будет муниципалитет [10].

---

[1] *Ивлева И.П.* Предпринимательская деятельность и понятие предпринимательского риска // Сборник материалов III Международной научно-практической конференции «Научно-технический прогресс: актуальные и перспективные направления будущего». Кемерово: Западно-Сибирский научный центр, 2016. С. 301–302.

[2] *Литвинова Е.А.* Государственная финансовая поддержка субъектов малого предпринимательства на муниципальном уровне, ее значение и мероприятия по повышению эффективности // Сборник материалов Всероссийской научно-практической конференции с международным участием «Проблемы устойчивого развития российских регионов». Тюмень: Тюменский индустриальный университет, 2014. С. 249–252.

[3] *Татарникова А.О.* Направления повышения эффективности использования государственных и муниципальных финансовых ресурсов в Российской Федерации // Сборник материалов X международной научно-практической конференции «Экономическая наука в 21 веке: вопросы теории и практики». Махачкала: Апробация, 2016. С. 84–85.

Выпуск переводных векселей может быть обеспечен будущими доходами бюджета, так как векселя к оплате «по предъявлению» можно принимать к оплате обязательств хозяйствующих субъектов перед бюджетом. Из средств бюджета, выделяемых на поддержку предпринимательства, целесообразно сформировать «Фонд инвестиций в микропредприятия».

При этом субъектам предпринимательской деятельности могут быть предоставлены микрокредиты по среднерыночной процентной ставке сроком на один год[4]. Предлагаемые условия кредитования позволяют минимизировать риски неисполнения обязательств заемщиками и максимизировать количество субъектов малого и среднего предпринимательства, получателей поддержки [11].

По данным кредитам целесообразно обеспечивать муниципальные гарантии собственными долговыми обязательствами в размере 10% от суммы кредита.

Рассмотрим предлагаемую схему кредитования на конкретном примере.

Субъект предпринимательства получает кредит в размере 500 000 руб.

Срок кредитования 1 год, ставка по кредиту 15% годовых.

Таким образом, общая сумма платежа по кредиту составит 575 тыс. руб.

Условия кредитования по данной схеме предполагают обеспечение поручительства или залога. Муниципалитет обеспечивает поручительство по данному кредиту в размере 10% от суммы кредита собственными переводными векселями на сумму 50 000 руб. Оставшуюся сумму в размере 450 000 руб. и проценты по кредитному договору в размере 75 000 руб. субъект предпринимательства обеспечивает своим поручительством или залогом с коэффициентом 1, что тоже можно рассматривать как льготное условие для заемщика, поскольку среднерыночные условия займов предполагают обеспечение залога с коэффициентом 1,5–1,6. То есть при обеспечении займа в 450 000 руб. на среднерыночных условиях стоимость залога должна составлять от 675 000 до 720 000 руб.[5] [12, 13].

В случае добросовестного исполнения обязательств «Фонд инвестиций в микропредприятия» получает процентный доход в сумме 75 000 руб. или 15% от инвестиций. А муниципалитет – назад свои векселя на сумму 50 000 руб. То есть средства бюджета, предназначенные на поддержку предпринимательства, не только остались в бюджете, но и могут быть использованы в данном направлении повторно. Поэтому доходность данной сделки для муниципалитета составит 16,7%.

В случае неисполнения обязательств заемщиком кредитный риск представляет собой сумму в размере 450 000 руб. суммы основного долга + 75 000 руб. процентов по кредиту, которые обеспечены залогом или поручительством заемщика, то есть максимальный финансовый риск заемщика определяется минимальной доходностью в размере 25 000 руб., или 5%.

Рассмотренный пример позволяет сделать вывод о том, что в целях облегчения доступа субъектов предпринимательства к финансовым ресурсам и минимизации кредитных рисков заемщика можно регулировать размер кредитной ставки так, чтобы риск (убыток) вложения был минимальным и определялся неравенством, в котором сумма размещенных средств по кредитным договорам под 15% годовых и 90-процентном исполнении обязательств заемщиком с учетом возможного невозврата от размещенных средств была меньше на 10% или равна сумме чистых инвестиций [14, 15].

---

[4] *Мадумаров Т.Т.* Основания и порядок предоставления микро-кредитов или микрозаймов // Сборник статей межвузовской научно-практической конференции с международным участием «Проблемы права в современной России». СПб.: Санкт-Петербургский политехнический университет Петра Великого, 2016. С. 140–147.

[5] *Морозова Ю.В.* Об оценке предметов залога и их реализации // Сборник статей Международной научно-практической конференции «Наука, образование и инновации». Уфа: ОМЕГА САЙНС, 2016. С. 127–130.

То есть если на рассмотренном примере снизить процентную ставку до 10%, то риск заемщика будет равен 0.

Кроме того, использование данной схемы кредитования дает возможность муниципалитетам финансировать предпринимательские проекты, реализуемые в приоритетных для экономики округов[6] [16] направлениях по еще более низким процентным ставкам, и принимать на себя 10-процентный риск, обеспеченный собственным поручительством.

Возможные риски неисполнения обязательств на 10% обеспечиваются средствами бюджета, опосредованными векселями, поэтому максимально возможный риск (убыток) не должен превышать 10% от инвестированных средств, то есть критическая доходность $E_{\text{Кр.}}$ должна быть не более –10%, минимальная доходность $E_{min} = 0\%$, а максимальная доходность инвестиций $E_{max}$ составляет 16,7% в случае 100% возврата кредитов.

При этом представляет интерес определение критических точек:

1) максимального процента невозвратов, при котором доходность «Фонда инвестиций в микропредприятия» будет равна 0.
$E_{min} = ((1-(1 \cdot 0{,}13)) \cdot 1{,}15 - 1)/1 \cdot 100 = 0\%$, то есть при невозврате 13% от вложенных средств риск (убыток) равен 0%;

2) критическая точка невозвратов, при которой доходность «Фонда инвестиций в микропредприятия» будет равна –10%, то есть для покрытия убытков не придется привлекать дополнительные бюджетные средства.

$E_{\text{Кр.}} = ((1-(1 \cdot 0{,}217)) \cdot 1{,}15 - 1)/1 \cdot 100 = -10\%$, то есть критическим значением невозврата является 21,7% от инвестиций, чтобы максимальный риск покрывался собственными долговыми обязательствами.

Таким образом, риск вложений можно считать минимальным, а эффективность вложений $E$ может колебаться в диапазоне [–10%; +16,7%] от минимально возможного значения (–10%) в случае 21,7% невозврата инвестиций, до максимального значения 16,7%, в случае 100% возврата денежных средств.

На основании изложенного можно утверждать, что предложенная методика позволяет определить при заданной норме доходности $K\%$ (ставке по кредиту):

1) максимальный процент невозвратов $X_0$, при котором доходность «Фонда инвестиций в микропредприятия» будет равна 0:

$E_{min} = ((1-(1 \cdot X_0)) \cdot K\% - 1) \cdot 100 = 0\%$,

$X_0 = 1 - (1/(1+K\%)) \cdot 100$;

2) критическую точку невозвратов, при которой доходность «Фонда инвестиций в микропредприятия» будет равна размеру поручительства ($РП$) бюджета по выданным кредитам:

$E_{\text{Кр.}} = ((1-(1 \cdot X_{\text{Кр.}})) \cdot K\% - 1) \cdot 100 = -РП\%$,

$X_{\text{Кр.}} = (1-(1-РП)/(1+K\%)) \cdot 100$.

Следовательно, при прогнозируемом риске невозврата денежных средств формирование «Фонда инвестиций в микропредприятия» и обеспечение части обязательств заемщика по кредиту государственными (муниципальными) долговыми ценными бумагами позволит расширить объемы финансирования и обеспечить доступ субъектов предпринимательства к финансовой поддержке на приемлемых и простых условиях как для заемщика, так и для кредитора.

---

[6] *Алябьева Е.М.* Исследование диспропорций регионального развития (на примере Самарской области) // Вестник Поволжского государственного университета сервиса. Серия: Экономика. 2017. № 1. С. 32–41.

16. *Лукоянчев С.С.* Предлагаемая схема оценки эффективности инвестиций // Современное развитие экономических и правовых отношений. Образование и образовательная деятельность. 2014. № 1. С. 275–279.

**Информация о конфликте интересов**

Я, автор данной статьи, со всей ответственностью заявляю о частичном и полном отсутствии фактического или потенциального конфликта интересов с какой бы то ни было третьей стороной, который может возникнуть вследствие публикации данной статьи. Настоящее заявление относится к проведению научной работы, сбору и обработке данных, написанию и подготовке статьи, принятию решения о публикации рукописи.

# THE REGULATION METHODS OF FISCAL RISK IN THE FRAMEWORK OF THE IMPLEMENTATION OF ENTREPRENEURSHIP SUPPORT

**Elena G. DMITRIK**

Stary Oskol Technological Institute, Branch of National University of Science and Technology (MISiS),
Stary Oskol, Russian Federation
dmitrikeg@mail.ru

**Article history:**
Received 31 July 2017
Received in revised form
14 August 2017
Accepted 30 August 2017
Available online
28 September 2017

**JEL classification:** G21, G28, H81

**Keywords:** financing, financial risks regulation, entrepreneurship

**Abstract**

**Importance** In this article, I consider the issues of financial resource shortage, the limited possibility of attracting bank loans, and business risk. All these problems constrain the development of entrepreneurship.
**Objectives** I aim to determine a scientifically based financial mechanism of financing the priority directions of territories' development.
**Methods** In this paper, I use economic and mathematical methods, logical analysis, and synthesis.
**Results** In this study, I develop tools of financing the priority directions of the municipal economy. The proposed financial scheme allows to expand the volume of financing and ensure the access of businesses to financial support.
**Relevance** The article proposes concrete financing mechanisms for investment with minimal risk for the budget and preferential conditions for business.

**Please cite this article as:** Dmitrik E.G. The Regulation Methods of Fiscal Risk in the Framework of the Implementation of Entrepreneurship Support. *Finance and Credit*, 2017, vol. 23, iss. 36, pp. 2189–2196.
https://doi.org/10.24891/fc.23.36.2189

## Acknowledgments

I express my deep gratitude and appreciation to Anna F. VINOKHODOVA, Doctor of Economics, Professor, for advice and valuable comments during the work on this article.

**Conflict-of-interest notification**

I, the author of this article, bindingly and explicitly declare of the partial and total lack of actual or potential conflict of interest with any other third party whatsoever, which may arise as a result of the publication of this article. This statement relates to the study, data collection and interpretation, writing and preparation of the article, and the decision to submit the manuscript for publication.